\documentclass[10pt,a5paper,twoside]{article}

\usepackage{graphicx}
\usepackage{fancyhdr}
\usepackage{amsmath}

\newcommand{\RECIVEDATE}{9 March 2008}
\newcommand{\ACCEPTEDDATE}{22 May 2008}
\newcommand{\VOL} {VI}
\newcommand{\NO} {1}
\newcommand{\FIRSTPAGE}{3}

\newcommand{\TITLE} {CONSERVED NOETHER CURRENTS, UTIYAMA'S THEORY OF INVARIANT VARIATION, AND VELOCITY DEPENDENCE IN LOCAL GAUGE INVARIANCE
}
\newcommand{\AUTHOR} {Gy\"orgy Darvas}
\newcommand{\SHORTAUTHOR} {\AUTHOR}
\newcommand{\SHORTTITLE}{Conserved Noether currents, Utiyama's theory ...}
\newcommand{\AFFILIATION} {
\small Symmetrion\\ 
\small and\\ 
\small Institute for Research Organisation, Hungarian Academy of Sciences \\
\small P.O.B. 994, Budapest, H-1245 Hungary \\
\small e-mail: darvasg@iif.hu
}

\makeatletter
\def\@hangfrom#1{%
    \setbox\@tempboxa\hbox{{#1}}%
    \hangindent \wd\@tempboxa%
    \noindent\box\@tempboxa}
\renewcommand\section{\@startsection {section}{1}{0cm}%
                                   {3.3ex plus 1ex minus 2ex}%
                                   {2.3ex plus 1ex minus 2ex}%
                                   {\normalfont\large\bfseries}}

\renewcommand\subsection{\@startsection {subsection}{2}{0cm}%
                                   {3.6ex plus 1ex minus 1ex}%
                                   {2.3ex plus 1ex minus 2ex}%
                                   {\normalfont\normalsize\bfseries}}

\renewcommand\subsubsection{\@startsection {subsubsection}{3}{0cm}%
                                   {3.6ex plus 1ex minus 1ex}%
                                   {2.3ex plus 1ex minus 2ex}%
                                   {\normalfont\normalsize\bfseries}}
\makeatother

\begin{document}

\pagestyle{fancy} \pagestyle{fancyplain} \thispagestyle{fancy}
\thispagestyle{fancyplain}

\fancyhf{} \cfoot{\small{\textsf{Concepts of Physics, Vol. \VOL, No.~\NO~(2009), pp. 3-16.
}}}
\renewcommand{\headrulewidth}{0pt}

\fancyfoot[LE,RO]{\thepage}

\begin{titlepage}

\begin{figure}[t]
\begin{center}
\end{center}
\end{figure}

\setcounter{page}{\FIRSTPAGE}

\title{\large\bf \TITLE}
\author{\small  \AUTHOR  \\
\AFFILIATION
}

\date{}
\maketitle \center{\textsl{\footnotesize (Received \RECIVEDATE; accepted \ACCEPTEDDATE)}}

\begin{abstract}
The paper discusses the mathematical consequences of the application of derived variables in gauge fields. Physics is aware of several phenomena, which depend first of all on velocities (like e.g., the force caused by charges moving in a magnetic field, or the Lorentz transformation). Applying the property of the second Noether theorem, that allowed generalised variables, this paper extends the article by Al-Kuwari and Taha (1991) with a new conclusion. They concluded that there are no extra conserved currents associated with local gauge invariance. We show, that in a more general case, there are further conserved Noether currents. In its method the paper reconstructs the clue introduced by Utiyama (1956, 1959) and followed by Al-Kuwari and Taha (1991) in the presence of a gauge field that depends on the co-ordinates of the velocity space. In this course we apply certain (but not full) analogies with Mills (1989). We show, that handling the space-time co-ordinates as implicit variables in the gauge field, reproduces the same results that have been derived in the configuration space (i.e., we do not lose information), while the proposed new treatment gives additional information extending those. The result is an extra conserved Noether current.
\end{abstract}

\pagestyle{fancy} \pagestyle{fancyplain} \thispagestyle{fancy}
\thispagestyle{fancyplain}

\fancyhf{} \cfoot{\small{\textsf{Concepts of Physics, Vol. \VOL, No.~\NO~(2009), pp. 3-16.}}}
\renewcommand{\headrulewidth}{0pt}

\fancyfoot[LE,RO]{\thepage}

\end{titlepage}

\pagestyle{fancy} \fancyhf{}
\renewcommand{\headrulewidth}{0pt}
\renewcommand{\sectionmark}[1]{\markright{\thesection\ #1}}
\cfoot{\small{\textsf{Concepts of Physics, Vol. \VOL, No.~\NO~(2009), pp. 3-16.}}}
\chead[\small{\textsf{\SHORTAUTHOR}}]{\small{\textsf{\SHORTTITLE}}}
\fancyfoot[LE,RO]{\thepage}

\section{Introduction}
Al-Kuwari and Taha (1991) discussed the conditions of local gauge invariance under a general non-Abelian group. They developed an interpretation of the Noether (1918) theorems applying the generalisation by Utiyama (1956, 1959), which imply the field equations for gauge vector fields and the existence of conserved Noether currents of global gauge invariance. They conclude that there are no extra conserved currents associated with local gauge invariance. We show, that in a more general case, there are further conserved Noether currents.

Noether's mathematical derivation allows the dependence of the concerned fields on any, general co-ordinate. So do Al-Kuwari and Taha when they apply the same, general parameters as dependent variables of physical fields, however, they do not discuss the consequences of the application of the possible variables other than the four space-time co-ordinates. In this sense they discuss the theorem in a restricted domain of variables. Such ignored, but allowed and possible variables are e.g., the co-ordinates in the velocity four-space. Indeed, certain fields may depend on the co-ordinates of the velocity space as well. In classical electrodynamics, e.g., the force caused by the current of moving electric charges depends on the velocity of the charges compared to the reference frame of the observer
$[F=\frac{\nu}{c}\times B ]$. In relativistic field theories the best known example for velocity dependent phenomena is the covariant effect of the Lorentz transformation [$(x^\mu)'=\Lambda_\nu^\mu(v)x^\nu$ for space-time vectors and $(F^{\mu\nu})'=\Lambda_\alpha^\mu(v)\Lambda_\beta^\nu(v)F^{\alpha\beta}$ for the electro-magnetic field tensor  (NB: summation convention on repeated indices is used throughout the document)]. Descriptions of the mentioned phenomena handle the space-time co-ordinates as indirect variables.

This paper reconstructs an apparent clue, by the cited authors, pertaining to the presence of a gauge field that depends on the co-ordinates of the velocity space (and imlicitly on the space-time co-ordinates). The result is an additional conserved Noether current that is localised in the velocity space.  For the effects of a general non-Abelian group on the described local gauge invariance we refer to R. Mills' (1989) review paper. We partially use the methods of his description of YM type gauge fields; however we introduce a new type of localised gauge field that does not coincide with the YM field. We use the same letter ({\bf D}) to denote this gauge field for other (here not detailed) considerations, but this {\bf D} field is {\em per definitionem} different from his.

Concerning the interpretation of localisation, we must emphasise (before one would mistakenly claim that there is a possible loss of causality in our theory) that we do not use the term localisation restricted to the placement in space and time, rather in a generalised meaning, when we extend the role of the co-ordinates to a set of generalised variables (cf. Noether, 1918). These variables may be the four space-time co-ordinates or they may be others (and their number may vary). In her mathematical terms of invariant variational problems, the space-time co-ordinates did not play a distinguished role. According to her second theorem, other variables, such as velocity-space co-ordinates, are allowed which may implicitly depend on the space-time co-ordinates. In general: $T[G_{\infty,\rho}]=T[p_\alpha(x_\beta)]$ $(\alpha=1,\dots,\rho)$, $(\beta=1,\dots,\sigma)$. {\em T} are the transformations that transform the elements of {\em G} into each other. The relation of the local transformations $V(\dot{x})\in G$, and the parameters {\em p}, are defined in Sec. 2, above the Eq. (1). The {\em p} are parameters on which the transformations, constituting the group elements, depend. They take the form of functions $p_\alpha(x_\beta)$ and their derivatives. The group transformations depend on {\em p} and are finitely differentiable.

First we will discuss the problem in the mathematical terms mentioned. Later we will discuss the possible physical consequences when we have derived the resulting equations. Anticipating possible objections concerning any loss of classical localisation, we draw attention to the fact that in our discussion, the {\bf D} gauge field introduced depends directly on the velocity-space co-ordinates, while the matter field depends directly on the 4D space-time co-ordinates. In other words, this means that although we primarily used co-ordinates of the velocity-space, our derivations are indirect and include derivatives with respect to the space-time co-ordinates (cf., the introduction of the $\lambda^\nu_\mu$ tensor) and play important role in our conclusions. This is an expression of the fact that we observe the physical events (occurring even in the velocity space) with respect to the 4D space-time.

One may now ask: Why do we take into consideration space-time co-ordinates in the form of implicit parameters (cf. also Note 2)? There are pragmatic reasons. Generally physics introduces dependence on the co-ordinates of a configuration space in the form $f(\dot{x}_\mu,x_\nu)$. In our case this treatment would lead to 8 parametric derivations, and large acceleration tensors, whose several elements would play no role in the discussion. For practical reasons we could reduce the calculations to the most necessary ones by replacing the $f(\dot{x}_\mu,x_\nu)$ dependence with a $f(\dot{x}_\mu(x_\nu))$ dependence. The localisation is present here too (in the above generalised, Noetherian sense), although we are allowed another way of calculating it.

We will consider Lagrangians which depend on matter fields $\phi_k$ and gauge fields $D_{\dot{\mu},\alpha}$, which all depend - in simple mathematical terms - on parameters. In physical terms these parameters are generally identified with the four space-time co-ordinates. Describing a specific situation: we have to define the concrete form of dependence of the gauge field on the space-time co-ordinates (in other terms, the concrete form of localisation). In our case this dependence of {\em D} on $x_\mu$  will be given by the formula: $D_{\dot{\mu}}=D_\mu\left(\frac{\partial x^\mu}{\partial x_4}\right)$,  or in another form $D_{\dot{\mu}}=D_\mu\left[\dot{x}^\mu(x_\nu)\right]$. The 2nd theorem of Noether is just about Lagrangians, which depend on arbitrary number of fields with arbitrary numbers of derivatives by arbitrary number of parameters. Why can we apply her theorem here? The reason is that in mathematical terms she did not specify either the physical-mathematical character or the number of applicable parameters.

One may be concerned that this could contradict the usual physical theory. We show that such anxiety does not hold. In a boundary situation, namely in the absence of a velocity-dependent gauge field - an identical situation to the one originally studied by Al-Kuwari and Taha (1991) - we obtain the same currents, that were derived by them in a space-time dependent field, cf. (21) and (23). In other words, in the absence of relativistically high velocities or acceleration, the effect of the velocity dependent gauge field can be neglected, and we get back to the same currents as derived in the former, cited work; cf. (23), and compare it with the (24) derived here. This boundary result justifies our choice and method. Nevertheless, in the presence of a velocity dependent gauge field, we derive new conserved Noether currents (22) and (24).

In short, the {\em novelty in our treatment} is two-fold: One is a merely mathematical-technical one, i.e., the implicit parametrisation of the space-time co-ordinates.  The other is the consideration of a gauge field {\bf D}, that depends on the temporal derivatives of the space-time co-ordinates; these may have physical meaning. The {\em novelty in our results} are the {\em new} conserved currents.

\section{Noether's currents for gauge invariance localised in the velocity space}

The presentation discusses general, non-Abelian case. Let's first introduce a {\bf D} field localised in the velocity space, whith components $D_{\dot{\mu}}= D_{\dot{\mu}}(\dot{x}^\mu)$\footnote{We should note that {\bf D} is defined in this paper with a different meaning, than in the later cited paper by R. Mills
(1989).}, where $\dot{x}^\mu=\dot{x}^\mu(x_\nu)$, $(\mu,\nu=1,2,3,4)$\footnote{Although this is a mathematical paper (cf., we do not use any concrete physical content of the applied Lagrangians, see the Note 3), one may ask whether there is any physical meaning of such a velocity-dependent field whose space-time dependence is considered only as defined here indirectly, or it is just a hypothetical mathematical construction. Nevertheless, the results obtained (cf., Sec. 3, Conclusions) justify the
assumption of this paper, that such a velocity-dependent field, may have physical meaning. According to the author's opinion, there are several physical situations, where such a description has relevance and pragmatic advantages. Let's give one example. 

Consider the following situation: We are taking measurements in a system of reference fixed to a lab. We are measuring effects of moving charges. The effects, e.g., force, originating from the moving reference system of the charges, as causes of the measured effect, depend in first approximation on three sets of parameters - their space and time coordinates, their velocity components, and the charge. If we fix the value of one of those sets, the effects measured by us will no longer directly depend on those. E.g., we can measure the effect originating from an electron. In this case the value of the charge is given and it will no longer play the role of an independent variable. Let's imagine the following situation: we are measuring effects of a valence electron oscillating in the electromagnetic field between two atoms whose position is fixed in a crystal. In this case, although the electron executes motion, the change in its space coordinates is neglectable small compared to the
distance from the measuring instrument in our lab. In practical terms its position does not change. The effect we are observing depends only on its actual velocity. Of course, we record its position in our lab, but we can consider this only as a dependent variable. We make use of this when we define the derivatives of the velocity in the reference system fixed to our lab.}  (Dotted indices denote the velocity-space components.) We will introduce a $\lambda^\nu_\mu$ tensor defined as $\lambda^\nu_\mu=\partial_\mu\dot{x}^\nu=\frac {\partial \dot{x}^\nu}{\partial x_\mu}$ which characterises the changes of the velocity-space components in the space-time. Localisation will be taken into consideration in this way (we refer to the generalised interpretation of localisation as defined above in Section 1).

Now, consider a Lagrangian density $L(\phi_k,D_{\dot{\mu},\alpha})$, where 
$\phi_k, k=1,\dots,n$ are the matter fields, which also includes the velocity 
field $\dot{x}^\mu=\dot{x}^\mu(x_\nu)$, and $D_{\dot{\mu},\alpha}, \alpha=1,
\dots,N$ are the gauge fields. We assume, that $L(\phi_k,D_{\dot{\mu},\alpha})$ 
is invariant under the local transformations of a compact, simple Lie group {\em G} 
generated by $T_{\alpha}, \alpha=1,\dots,N$ where 
$[T_\alpha,T_\beta]=iC_{\alpha\beta}^\gamma T_\gamma$ where 
$C_{\alpha\beta}^\gamma$ are the so-called structure constants. 
For simplicity we assume, similar to Al-Kuwari and Taha, that the 
matter fields belong to a single, {\em n}-dimensional representation 
of {\em G}. Under a local transformation $V(\dot{x})\in G$ parameterised 
by $p_\alpha(\dot{x})$   
\begin{equation*}
V(\dot{x})=e^{-ip_\alpha(\dot{x})T_\alpha}
\end{equation*}
the infinitesimal transformations of the matter- and the gauge fields can be formulated as:
\begin{equation}
\delta\phi_k=-ip_\alpha(\dot{x})(T_\alpha)_{kl}\phi_l(\dot{x})\qquad k=1,\dots,n,
\end{equation}
where the $T_\alpha$ are matrix-representation operators generating the group {\em G}, with the above commutation rule $[T_\alpha,T_\beta]=iC_{\alpha\beta}^\gamma T_\gamma$,
and
\begin{equation}
\delta D_{\mu,\alpha}=\frac{1}{g}\partial^{\dot{\rho}} p_\alpha (\dot{x})\partial_\mu \dot{x}^\rho + C_{\alpha\beta}^\gamma p_\beta (\dot{x}) D_{\dot{\mu},\gamma}(\dot{x})\qquad \alpha=1,\dots,N
\end{equation}
where $\partial^{\dot{\rho}}=\frac{\partial}{\partial\dot{x}^\rho}$.

For the induced infinitesimal transformation $\delta L$ of the Lagrangian density $L(\phi_k,D_{\dot{\mu},\alpha})$, on using the field equations for both the matter and the gauge fields, one obtains
\begin{equation}
\delta L= \partial_\mu \left[\frac{\partial L}{\partial (\partial_\mu\phi_k)}\delta\phi_k+ \frac{\partial L}{\partial (\partial_\mu D_{\nu,\alpha})}\delta D_{\nu,\alpha} \right]
\end{equation}
One would like to describe the events, recorded in the velocity-space dependent gauge field, as they are observed from the 4D space-time. Therefore one needs to apply derivatives by the space-time co-ordinates. Substituting from (1) and (2) into (3), using the notation $\frac{\partial\dot{x}^\nu}{\partial x_\mu}=\partial_\mu\dot{x}^\nu=\lambda^\nu_\mu$ and a permutation of the indices, one can obtain
\begin{equation*}
\begin{split}
\delta L &= \partial_\mu \left[\frac{\partial L}
{\partial (\partial_\mu\phi_k)}(-i)p_\alpha(\dot{x})(T_\alpha)_{kl}\phi_l(\dot{x})\right]
+\\
&+ \partial_\mu\left[ \frac{\partial L}{\partial (\partial_\mu D_{\dot{\nu},\alpha})} 
\frac{1}{g}\partial^{\dot{\rho}} p_\alpha (\dot{x}) \lambda^\rho_\nu \right]
+ \partial_\mu\left[\frac{\partial L}{\partial (\partial_\mu D_{\dot{\nu},\alpha})} C_{\alpha\beta}^\gamma p_\beta (\dot{x}) D_{\dot{\nu},\gamma}(\dot{x}) \right]
\end{split}
\end{equation*}
and from this
\begin{equation}
\begin{split}
\delta L &= \partial_\mu \left[\frac{\partial L}{\partial (\partial_\mu\phi_k) }\frac{1}{i} (T_\beta)_{kl}\phi_l(\dot{x})- C_{\alpha\beta}^\gamma \frac{\partial L}{\partial (\partial_\mu D_{\dot{\nu},\alpha})} D_{\dot{\nu},\gamma}(\dot{x}) \right] p_\beta(\dot{x}) +\\
&+\left[- \frac{\partial L}{\partial (\partial_\mu\phi_k) }\frac{1}{i} (T_\beta)_{kl}\phi_l(\dot{x})\lambda^\nu_\mu +
C_{\alpha\beta}^\gamma \frac{\partial L}{\partial (\partial_\mu D_{\dot{\rho},\alpha})} D_{\dot{\rho},\gamma}(\dot{x})\lambda^\nu_\mu -\right.\\
&-\left. \frac{1}{g} \frac{\partial L}{\partial (\partial_\rho D_{\dot{\mu},\beta})}\partial_\rho \lambda^\nu_\mu
- \frac{1}{g} \partial_\rho(\frac{\partial L}{\partial (\partial_\rho D_{\dot{\mu},\beta})}) \lambda^\nu_\mu
\right] \partial^{\dot{\nu}} p_\beta(\dot{x}) -\\
&-\frac{1}{g}(\frac{\partial L}{\partial (\partial_\mu D_{\dot{\nu},\alpha})}) \lambda^\rho_\nu\partial_\mu \partial^{\dot{\rho}} p_\alpha(\dot{x})
\end{split}
\end{equation}
Al-Kuwari and Taha observed, that  
$p_\alpha(\dot{x})$, $\partial^{\dot{\mu}} p_\alpha(\dot{x})$, and  
$\partial_\mu\partial^{\dot{\nu}}p_\alpha(\dot{x})$, $\alpha=1,\dots,N$ 
are arbitrary and independent at any  $\dot{x}$, so that the local gauge invariance of  
$L(\phi_k,D_{\dot{\mu},\alpha})$, $\delta L=0$ is equivalent to the following three 
conditions:
\begin{equation}  \label{cond_one}
\begin{split}
\partial_\mu \left[\frac{\partial L}{\partial (\partial_\mu\phi_k) }\frac{1}{i} (T_\alpha)_{kl}\phi_l(\dot{x})- C_{\alpha\beta}^\gamma \frac{\partial L}{\partial (\partial_\mu D_{\dot{\nu},\beta})} D_{\dot{\nu},\gamma}(\dot{x}) \right] =0\end{split}
\end{equation}

\begin{subequations}
\begin{equation}  
\begin{split}
\frac{\partial L}{\partial (\partial_\mu\phi_k)}\frac{1}{i} 
(T_\alpha)_{kl}\phi_l(\dot{x})\lambda^\nu_\mu &-
C_{\alpha\beta}^\gamma \frac{\partial L}{\partial (\partial_\mu D_{\dot{\rho},\beta})} 
D_{\dot{\rho},\gamma}(\dot{x})\lambda^\nu_\mu +\\
&+ \frac{1}{g} \partial_\rho\left[\frac{\partial L}
{\partial (\partial_\rho D_{\dot{\mu},\alpha})} \lambda^\nu_\mu
\right] =0
\end{split}
\end{equation}

or written in another form:
\begin{equation} \label{cond_2b}
\begin{split}
\frac{\partial L}{\partial (\partial_\mu\phi_k)}\frac{1}{i} 
(T_\alpha)_{kl}\phi_l(\dot{x})\lambda^\nu_\mu &-
C_{\alpha\beta}^\gamma \frac{\partial L}{\partial (\partial_\mu D_{\dot{\rho},\beta})} 
D_{\dot{\rho},\gamma}(\dot{x})\lambda^\nu_\mu +\\
+ \frac{1}{g} \frac{\partial L}{\partial (\partial_\rho D_{\dot{\mu},\alpha})} 
\partial_\rho\lambda^\nu_\mu
&+ \frac{1}{g} \partial_\rho\left[\frac{\partial L}
{\partial (\partial_\rho D_{\dot{\mu},\alpha})}\right] \lambda^\nu_\mu =0
\end{split}
\end{equation}
\end{subequations}

\begin{subequations}
\begin{equation}  \label{cond_3a}
\frac{\partial L}{\partial (\partial_\mu D_{\dot{\nu},\alpha})} \lambda^\rho_\nu=0
\end{equation}
or considering that $\partial_\mu \partial^{\dot{\rho}} p_\alpha(\dot{x})$ is symmetric in $\mu$ and $\rho$  :
\begin{equation}
\frac{\partial L}{\partial (\partial_\mu D_{\dot{\nu},\alpha})} \lambda^\rho_\nu+
\frac{\partial L}{\partial (\partial_\nu D_{\dot{\mu},\alpha})} \lambda^\rho_\mu = 0
\end{equation}
\end{subequations}

Let's define
\begin{equation}
F_\alpha^{\mu\nu}(\dot{x})=
\frac{\partial L}{\partial (\partial_\mu D_{\dot{\rho},\alpha})} \lambda^\nu_\rho
\end{equation}
and
\begin{equation}
J_\alpha^{(1)\mu}=g\left[
i\frac{\partial L}{\partial (\partial_\mu\phi_k) }(T_\alpha)_{kl}\phi_l(\dot{x})+ 
C_{\alpha\beta}^\gamma \frac{\partial L}{\partial (\partial_\mu D_{\dot{\nu },\beta})} 
D_{\dot{\nu },\gamma}(\dot{x})
\right]
\end{equation}

\begin{equation}
\begin{split}
j_\alpha^{(2)\nu }=g\left[
i\frac{\partial L}{\partial (\partial_\mu\phi_k) } 
(T_\alpha)_{kl}\phi_l(\dot{x})\lambda^\nu_\mu \right.&+
C_{\alpha\beta}^\gamma \frac{\partial L}{\partial (\partial_\mu D_{\dot{\rho},\beta})} 
D_{\dot{\rho},\gamma}(\dot{x})\lambda^\nu_\mu - \\
&-\left. \frac{1}{g} \frac{\partial L}{\partial (\partial_\rho D_{\dot{\mu},\alpha})} 
\partial_\rho\lambda^\nu_\mu
\right]
\end{split}
\end{equation}

\medskip
\noindent
({\bf a}) Using (5), (6b) as well as (7a) and taking into consideration the condition that all elements of the tensor 
$\lambda^\nu_\mu$ are not zero, one obtains:
\begin{equation}
j_\alpha^{(2)\nu}(\dot{x})=J_\alpha^{(1)\mu}(\dot{x})\lambda^\nu_\mu
\end{equation}

One can observe that (11) mixes the components of the gauge-field currents 
$J_\alpha^{(1)\mu}$  and $j_\alpha^{(2)\nu}$  depending on the 4D velocity 
space in a similar way, like the Lorentz transformation mixes the co-ordinates 
of four-vectors in the 4D space-time. Note, that the $\lambda^\nu_\mu$  tensor 
characterises the changes of the velocity-space components in the space-time.

Now, let's investigate, what conditions the two currents, namely  $J_\alpha^{(1)\mu}$  
and $j_\alpha^{(2)\nu}$  fulfil, and if there are any, then which invariance conditions:

\medskip
\noindent ({\bf b}) Substituting  $J_\alpha^{(1)\mu}$ in (5), as well as in (6a) taking 
into consideration (7a) with the condition that all elements of the tensor 
$\lambda^\nu_\mu$ are not zero, one obtains:

\begin{equation}
\partial_\nu J_\alpha^{(1)\nu} =0
\end{equation}
and
\begin{equation*}
J_\alpha^{(1)\nu} = \partial_\rho \left(\frac{\partial L}{\partial (\partial_\rho D_{\dot{\mu},\alpha})}\right) 
\end{equation*}
Defining
\begin{equation*}
\frac{\partial L}{\partial (\partial_\mu D_{\dot{\nu},\alpha})}=F_\alpha^{(1)\mu\nu}
\end{equation*}
\begin{equation}
J_\alpha^{(1)\nu}  = \partial_\mu F_\alpha^{(1)\mu\nu}
\end{equation}
(12) and (13) - together with (7b) - means, that $J_\alpha^{(1)\nu}$ is a conserved 
Noether current.

\medskip
\noindent ({\bf c}) Substituting  $j_\alpha^{(2)\nu}$ in (6b) and taking again into consideration (7a) with the condition that all elements of the tensor $\lambda^\nu_\mu$ are not zero, one can obtain, that
\begin{equation}
j_\alpha^{(2)\nu} = \partial_\mu F_\alpha^{\mu\nu}
\end{equation}
The divergence of $j_\alpha^{(2)\nu}$:   $\partial_\nu j_\alpha^{(2)\nu}= \partial_\nu\partial_\mu F_\alpha^{\mu\nu}$  
does not vanish identically. Let's define $\alpha=1,\dots,N$ contravariant forms  $F_\alpha^{(2)\mu\nu}$-s so that their covariant derivatives be equal to the derivatives of $F_\alpha^{\mu\nu}$-s:
\begin{equation}
\hat{\partial}_\mu F_\alpha^{(2)\mu\nu}(\dot{x}) = \partial_\mu F_\alpha^{\mu\nu}(\dot{x})
\end{equation}
where the covariant derivative of $F_\alpha^{(2)\mu\nu}$  can be written 
as\footnote{Note the following: The YM theory (Yang and Mills, 1954; Mills, 1989) 
introduced the covariant form of $F_\alpha^{(2)\mu\nu}$ derived from the Lagrangian 
density of a specific fermion field. We do not make any preliminary assumption 
concerning the Lagrangian density of the field. This paper defines the covariant 
$F_\alpha^{(2)\mu\nu}$ in an essentially different, independent way, based on the 
requirement of their invariant transformation, and thus gets rid of any specific 
form of the Lagrangian density. The importance of this different approach becomes 
apparent looking at the discussion of the results by R. Mills himself (1989) in the 
light of the theory of fiber bundles. He observes that the applied covariant derivatives 
bear a very close relationship to the covariant derivatives of general relativity theory; 
and the quantities $F_\alpha^{(2)\mu\nu}$ are in close analogy to the curvature tensor 
of general relativity. Since YM theory derived $F_\alpha^{(2)\mu\nu}$ from a specific 
form for the Lagrangian density, they could not state anything more than an observed 
similarity. Furthermore, the Lagrangian-invariant introduction of $F_\alpha^{(2)\mu\nu}$ 
and their covariant derivatives also leaves free the opportunity for application to gravitational fields. An advantage of
this treatment is to find conserved Noether currents which are of an identical-form in different gauge fields.}
\begin{equation}
\hat{\partial}_\mu F_\alpha^{(2)\mu\nu}(\dot{x}) = \partial_\mu F_\alpha^{(2)\mu\nu} + igC_{\alpha\beta}^\gamma D_{\dot{\omega},\beta}\lambda^\omega_\mu \times F_\gamma^{(2)\mu\nu}
\end{equation}

\noindent As introduced above, the $T_\gamma$ (which appear in the currents) are matrix-representation operators generating the group {\em G}, with the already-mentioned commutation rule $[T_\alpha,T_\beta]=i C_{\alpha\beta}^\gamma T_\gamma$. In the case of {\em SU}(2) symmetry, {\em G} consists of $2\times 2$ matrices with 3 independent components, representing a state doublet, and in the case of {\em SU}(3) its matrix has 8 independent components, representing a state triplet.

The following $F_\alpha^{(2)\mu\nu}$ fulfil the requirement formulated in (16):
\begin{equation}
F_\alpha^{(2)\mu\nu}(\dot{x}) = \frac{\partial D_{\dot{\rho},\alpha}\lambda^\rho_\mu}{\partial x_\nu} - \frac{\partial D_{\dot{\sigma},\alpha}\lambda^\sigma_\nu}{\partial x_\mu} -igC_{\alpha\beta}^\gamma D_{\dot{\rho},\beta}\lambda^\rho_\mu D_{\dot{\sigma},\gamma}\lambda^\sigma_\nu
\end{equation}
This covariant $F_\alpha^{(2)\mu\nu}$  transforms under a non-Abelian, velocity-dependent gauge transformation in the same way as the isovector $F_\alpha^{\mu\nu}$. Thus one can replace the divergence of $F_\alpha^{\mu\nu}$  by the divergence of  $F_\alpha^{(2)\mu\nu}$.

Using (16) then (14)
\begin{equation}
\partial_\mu F_\alpha^{(2)\mu\nu} = \partial_\mu F_\alpha^{\mu\nu} - igC_{\alpha\beta}^\gamma D_{\dot{\omega},\beta}\lambda^{\omega}_{\mu} \times F_\gamma^{(2)\mu\nu}\\
\end{equation}
\begin{equation*}
\partial_\mu F_\alpha^{(2)\mu\nu} = j_\alpha^{(2)\nu} - igC_{\alpha\beta}^\gamma D_{\dot{\omega},\beta}\lambda^\omega_\mu \times F_\gamma^{(2)\mu\nu}
\end{equation*}
(18) defines a current
\begin{eqnarray}
J_\alpha^{(2)\nu} &= \partial_\mu F_\alpha^{\mu\nu} - igC_{\alpha\beta}^\gamma D_{\dot{\omega},\beta}\lambda^\omega_\mu \times F_\gamma^{(2)\mu\nu}\\
J_\alpha^{(2)\nu}  &= \partial_\mu F_\alpha^{(2)\mu\nu}(\dot{x})
\end{eqnarray}
whose four-dimensional divergence is automatically zero, due to (15) and (16).

\section{Conclusions}

Referring to (12), (13) and (15), (16), (20) we have obtained the following two sets of equations:
\begin{equation}
J_\alpha^{(1)\nu} = \partial_\mu F_\alpha^{(1)\mu\nu} \qquad \partial_\nu J_\alpha^{(1)\nu}=0
\end{equation}
\begin{equation}
J_\alpha^{(2)\nu} = \partial_\mu F_\alpha^{(2)\mu\nu} \qquad \partial_\nu J_\alpha^{(2)\nu}=0
\end{equation}

Completed with (7b) this set demonstrates, that in the presence of a velocity-dependent gauge field, there exist two (families of) conserved Noether currents.

In the absence of the velocity-dependent gauge field we get back to the results concluded by Al-Kuwari and Taha (1991) based on calculations in space-time dependent gauge field. Thus, without employing accelerations, we derived the same conserved currents. This justifies our preliminary assumption, that handling the space-time co-ordinates as implicit parameters not only provides additional information, but it preserves the physical relevance of the theory.

Although the two conserved currents are not independent, in the presence of a velocity-dependent gauge field they exist simultaneously. But which of the currents has the dominant effect will depend on the velocity of the described objects relative to the reference frame of the observer.

Taking into account (5) and (7a), one can now write (9) as
\begin{equation}
J_\alpha^{(1)\nu} = ig\frac{\partial L}{\partial(\partial_\nu\phi_k)}(T_\alpha)_{kl}\phi_l(\dot{x})
\end{equation}

The most significant conclusion of this paper is that in the presence of a velocity-dependent gauge field {\bf D}, there appear extra $J_\alpha^{(2)\nu}$  conserved currents. Taking into account (6b), (10) and (11), then (7a), one can now write (19) in the form
\begin{equation}
J_\alpha^{(2)\nu} = ig\left[\frac{\partial L}{\partial(\partial_\mu\phi_k)}(T_\alpha)_{kl}\phi_l(\dot{x})\lambda_\mu^\nu-
C_{\alpha\beta}^\gamma D_{\dot{\omega},\beta}\lambda^\omega_\mu\times F_\gamma^{(2)\mu\nu}
\right]
\end{equation}

Their dependence on the velocity-space gauge is apparent, although, none of the conserved vector currents involve the gauge parameters $p_\alpha(\dot{x})$ . From (24) with (23) [and (11), (15)] we see that $J_\alpha^{(1)\nu}$ and $J_\alpha^{(2)\nu}$ are not independent, yet they exist simultaneously.

From (21) and (23), considering (13), one obtains
\begin{equation}
\partial_\mu F_\alpha^{(1)\mu\nu} = ig\frac{\partial L}{\partial(\partial_\nu\phi_k)}(T_\alpha)_{kl}\phi_l(\dot{x})
\end{equation}
From (22) and (24), considering (18) and (16) one obtains
\begin{equation}
\hat{\partial}_\mu F_\alpha^{(2)\mu\nu} = ig\frac{\partial L}{\partial(\partial_\mu\phi_k)}(T_\alpha)_{kl}\phi_l(\dot{x})\lambda_\mu^\nu
\end{equation}

Relations (25) and (26) provide the equations of motion for the potential part\footnote{I.e., which serves as the source for the gauge-fields, and consequently as the source for the characteristic charges of the given fields.}  of the system's Lagrangian density. According to a note by Brading and Brown (2000), based on an observation of Utiyama (1959, p. 27), it is generally the case that when (25) or (26) is satisfied, the matter-field current associated with the Lagrangian acts as the source for the gauge fields. The note (Brading and Brown, 2000; Utiyama, 1959) is a consequence of the fact that the matter-field dependent and the gauge-field dependent currents are separate sides in each of the latter two equations\footnote{Brading and Brown (2000) call the equations defined in (21) - (26) ``coupled field equations'', after the form of the
connection that they describe between the different (matter- and gauge-) fields appearing in the Lagrangian. This holds always when the Euler-Lagrange equations are assumed to be satisfied for all the fields on which the Lagrangian
depends (or more precisely, for all the fields whose transformations depend on $\partial^{\dot{\mu}} p_\alpha$ ).}.  Here the only condition assumed was that the field equations be satisfied. No restriction was imposed on the form of the Lagrangian density except that it be invariant under local gauge transformations as defined in (3). The covariant dependence on the velocity-space gauge field is obvious from (26).

\section{Acknowledgement}
The author expresses his many thanks to the memory of {\em Yuval Ne'eman} for his wise advice in the course of personal consultations and correspondence during the years when the idea of this paper was shaped step by step. He is also indebted to {\em Laurence I. Gould} for suggestions during the preparation of this paper.


\end{document}